\documentclass[twocolumn,showpacs,aps]{revtex4}

\usepackage{graphicx}
\usepackage{dcolumn}
\usepackage{bm}

\begin{document}

\begin{titlepage}

\title{Ferroelectricity in Pb(Zr$_{0.5}$Ti$_{0.5}$)O$_{3}$ thin
films: critical thickness and 180$^\circ$ stripe domains}
\author{Zhongqing Wu, Ningdong Huang, Zhirong Liu, Jian Wu,
Wenhui Duan\footnote{Author to whom any correspondence should be
addressed}, Bing-Lin Gu} \affiliation{Center for Advanced Study
and Department of Physics, Tsinghua University, Beijing 100084,
People's Republic of China}
\author{Xiao-Wen Zhang}
\affiliation{State Key Laboratory of New Ceramics and Fine
Processing, Department of Materials Science and Engineering,
Tsinghua University, Beijing 100084, People's Republic of China}
\date{\today}

\begin{abstract}
The ferroelectric properties of disorder
Pb(Zr$_{0.5}$Ti$_{0.5}$)O$_{3}$ thin films are investigated with
Monte Carlo simulations on the basis of a first-principles-derived
Hamitonian. It is found that there exists a critical thickness of
about three unit cells ($\sim$12 \AA) below which the
ferroelectricity disappears under the condition that the in-plane
polarizations are suppressed by sufficient clamping effect. Above
the critical thickness, periodic $180^{\circ}$ stripe domains with
out-of-plane polarizations are formed in the systems in order to
minimize the energy of the depolarizaing field. The stripe period
increases with increasing film thickness. The microscopic
mechanism responsible for these phenomena is discussed.

\end{abstract}

 \maketitle

\draft

\vspace{2mm}

\end{titlepage}

During the past decade, ferroelectric thin films have been
investigated with a great deal of interest not only because their
physical properties are different from those of bulk
materials\cite{bra} but also because they are promising for
microelectronic and micromechanical applications\cite{sco}. A
fundamental problem attracting considerable attentions for
ferroelectric thin film is the critical thickness for the
ferroelectricity to occur. The observations that the ferroelectric
ground states exist in 40-\AA-thick perovskite oxide film
\cite{tyb} and in 10-\AA-thick crystalline copolymer\cite{bun}
seem to suggest the absence of the critical thickness. Some
theoretical investigations supported that opinion by predicting
ferroelectric ground states in various perovskite
slabs\cite{gho,mey}. Recently, Junquera and Ghosez investigated a
realistic ferroelectric-electrode interface and, contrary to the
current thought, revealed a critical thickness of about six unit
cells in BaTiO$_3$ thin film\cite{jav}. It was proposed that the
depolarizing electrostatic field\cite{daw} is responsible for the
disappearance of the ferroelectricity. However, in their study,
only the monodomain case, where the atomic off-center
displacements in every cell are the same, was considered. In fact,
polydomains or other atomic off-center displacements may come into
being to remove the depolarizing electrostatic field. For example,
Fu and Bellaiche found that a size reduction can lead to an
unusual atomic off-center displacements vortex pattern in
BaTiO$_3$ quantum dots \cite{fu}. Therefore, it is worth
investigating whether the film will form special atomic off-center
displacements pattern to remove the depolarization field and
whether the nonzero critical thickness still exists in that case.

In this paper, by use of a first-principles-based approach, the
ferroelectricity of disorder Pb(Zr$_{0.5}$Ti$_{0.5}$)O$_3$ thin
films is investigated to clarify the role of the strain and the
off-center displacement pattern in determining the critical
thickness. It is shown that the ferroelectricity in thin films is
closely related to the strain constraint imposed by the substrate.
When the thin films are free from constraint, a nonzero
polarization with an in-plane direction always exists in the
system, which does not disappear even in the monolayer films. When
there is a  compressive strain, the in-plane polarization is
suppressed while the out-of-plane polarization exhibits a strong
dependence on the film thickness. Above a critical thickness of
about three unit cells, the out-of-plane polarization forms
periodic $180^{\circ}$ stripe domains, and the period increases
with the film thickness. Below the critical thickness, the stripe
domain structure disappears, which suggests that the
ferroelectricity can be suppressed even in the absence of the
depolarizing electrostatic field. The orientation dependence of
dipole-dipole interaction is revealed to be responsible for these
phenomena, which is expected to generally work in perovskite
ferroelectric films.

We adopt the effective Hamiltonian of PZT alloys proposed by
Bellaiche, Garcia and Vanderbilt \cite{LB1,LB2,WZ} to predict the properties of the
disorder Pb(Zr$_{0.5}$Ti$_{0.5}$)O$_3$ thin film surrounded by the
vacuum.  All the parameters of the
Hamiltonian are derived from the first principle calculations and
are listed in references \cite{LB1,LB2}. Cohen \cite{coh} and Meyer \textit{et al.}\cite{mey1}
demonstrated that the effect of surface relaxation is significant in perovskites.
However, it should be noted that the effect of surface relaxation has been considered
implicitly in our simulations \cite{note1}, and  the influence of the surface upon the
ferroelectric order parameter is modest, as pointed out by Meyer \textit{et al.} \cite{mey1}.
In our simulations, we do not include an external term
of surface effect proposed by Fu and Bellaiche while simulating nanoscopic
structures, since they demonstrated that the term has almost no
effect on the polarization pattern\cite{fu}. From the above effective Hamiltonian,
Monte Carlo simulations are conducted for large supercells
typically containing between 5000 and 50000 atoms mimicking the
studied structures. The supercell average
 of the local soft modes $\mathbf{u}_{i}$ ($i$ is the cell index) is directly
proportional to the macroscopic electrical polarization. The
influence of the substrate is imposed by confining the homogeneous
in-plane strain.

To efficiently calculate the long-range dipole-dipole interaction
energy in thin films which lack the periodicity in the
out-of-plane direction, we adopt the corrected three-dimensional
Eward method, whose validity has been verified analytically by
Br\'{o}dka and Grzybowski \cite{bro}. In that scheme, a small
empty space (about three times of the film thickness) introduced
in the simulation box to surround the film can lead to very well
converged results. It is noted that the Eward sum we adopted for the
macroscopic sphere is immersed in vacuum. Therefore the effect of
depolarization field is considered in our model.

In our simulation, the $z$ axis ([001] direction) lies along the
growth direction of the film, and the $x$ and $y$  axes are chosen
to be along the pseudocubic [100] and [010]  directions. The film
thickness $d$ is measured by the number of unit cells along the
$z$-axis. As the first step, we simulate the ferroelectric
properties of thin films without any constrain from the substrate,
i.e., the free thin films are considered. It is found that for all
the film thicknesses considered, there always exists a nonzero
polarization along the $x$- or $y$- axis, and correspondingly, a
tensile strain of about 2\% along the same direction. (Data is not
shown here.) It is consistent with the simulation results in
quantum dots\cite{fu} where the normal component of polarization
near the surface is suppressed by the vacuum. It also verifies the
absence of any critical thickness for ferroelectricity in thin
films under the condition of vanishing internal stress\cite{mey}.

In actual systems, the film strain will be imposed by the
substrate (clamping effect) which has important effects on the
ferroelectricity\cite{per,osw,zhang,bal}. So we consider the
strain effects in our simulations. We calculate the supercell
average ($\overline{u}_x$, $\overline{u}_y$, $\overline{u}_z$) of
the local soft mode as functions of the in-plane strain. The
result with $d=4$ is shown in Fig.~1. It is clearly seen that
$\overline{u}_x$ and $\overline{u}_y$ decrease with increasing
(compressive) strain and disappear after the strain reaches a
critical value, while $\overline{u}_z$ keeps zero at any strain.
The disappearance of the in-plane polarization is of the first
order. It should be noted that when the in-plane strain is fixed,
$\overline{u}_x$ is equal to $\overline{u}_y$ due to the symmetry
requirement. Small difference between $\overline{u}_x$ and
$\overline{u}_y$ results from our treatment of avoiding effects of
symmetry-equivalent rotations of the order parameter\cite{WZ}.
Detail simulations indicate that the critical strain ($\sim 2\%$)
is insensitive to the film thickness. From the above results, it
can be concluded that the ferroelectricity does not disappear in
thin films (even for the monolayer case) if the strain imposed by
the substrate is not strong enough to suppress the in-plane
polarization.

\begin{figure}[tbp]
\includegraphics[width=8.5cm]{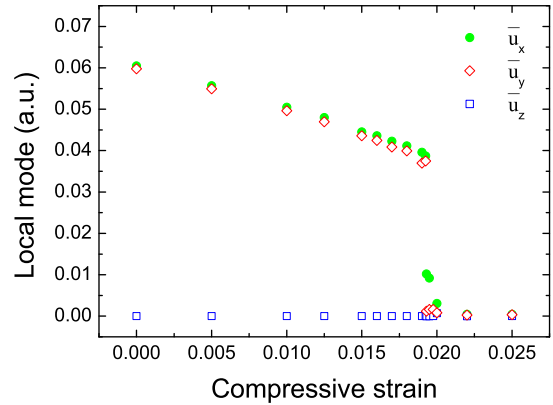}
\caption{The dependence of the supercell average ($\overline{u}_x,
\overline{u}_y$, $\overline{u}_z$) of the local soft mode on the
strain for the film with the thickness of four unit cells (i.e.,
$d=4$)}. \label{fig02}
\end{figure}

Although the average value of the out-of-plane polarization
($\overline{u}_z$) always keeps zero, the local mode $u_z$
exhibits completely different behavior below and above the
critical strain. $u_z$ is substantially small in any unit cell
below the critical strain, while it quickly increases nearby the
critical strain. Above the critical strain, periodic 180$^\circ$
stripe domains are formed in the system. An example with stripe
domains aligned along $y$-axis is shown in Fig.~2 for a thin film
with $d=4$ under 2\% compressive strain. The domain walls are
rather diffuse and the spatial variation of $u_z$ is close to a
cosine function. The similar stripe domain stabilized by strain
has also been found by Tinte \textit{et al.}\cite{Tin}. In the
general cases, we find that the direction along which the stripe
domains align is not certain. It may be the $x$ or $y$ axis and
may also have an angle with the $x$ or $y$ axis. In each stripe
domain, the $z$ component ($u_z$) of local soft modes have the
same sign. In other words, the stripe structure extends through
the film in the $z$ direction, which is important for
ferroelectric field effect and has been observed in PZT\cite{ahn}
and PbTiO$_3$ (PT) film \cite{str}. It is noted that the stripe
domain only appears above a certain film thickness. We calculate
many kinds of supercells with $d < 4$. No domain structure or
ferroelectricity is found in these supercells at 50 K even when
the compressive strain is as large as 4$\%$. Therefore, there is a
critical thickness of about three unit cells ($\sim$1.2 nm) in
disorder Pb(Zr$_{0.5}$Ti$_{0.5}$)O$_3$ thin films below which the
ferroelectricity disappears at large strains ($>2\%$).

\begin{figure}[tbp]
\includegraphics[width=8.5cm]{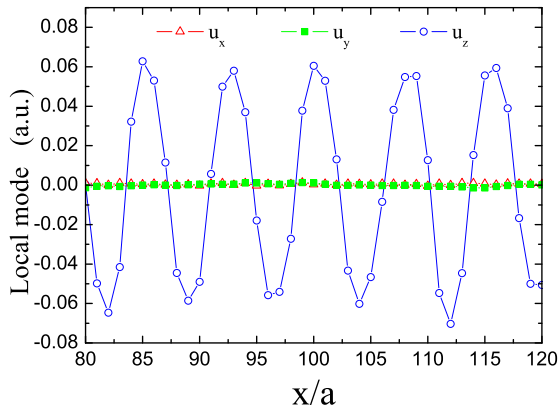}
\caption{The $y$-$z$ plane average ($u_x, u_y$, and $u_z$) of the
local soft mode as a function of $x/a$ ($a$ is the lattice
constant) for $120 \times 5\times 4$ supercell under 2$\%$
compressive strain. Only the data of $80<x<120$ are shown.  }
\end{figure}

The dependence of ferroelectric-paraelectric phase transition
temperature $T_c$ on the film thickness also supports the
existence of the critical thickness, as shown in Fig.~3. The
temperature is rescaled as in Ref. \cite{gar} due to the fact that
the effective-Hamiltonian approach overestimates $T_c$. We find
that $T_c$, determined by the temperature dependence of the
polarization amplitude (Fig.~3 inset), is far below $T_c$ of the
bulk material (dash line in Fig.~3). This is consistent with the
experimental observation of a significant decrease of $T_c$ in
thinner films (below 20 nm)\cite{str}. As the film thickness
decreases, $T_c$ decreases quickly. Extrapolation of $T_c$ in
Fig.~3 shows that the ferroelectricity disappears at the thickness
of about $d=3$. It is interesting to note that Ghosez and Rabe
also observed a critical thickness of 3 layers in PbTiO$_3$ with
completely different electrical boundary condition (perfectly
screened condition)\cite{gho}. Then, it seems that the electrical
boundary conditions have little effect on the critical thickness,
which is contrary to the common view.

\begin{figure}[tbp]
\includegraphics[width=8.5cm]{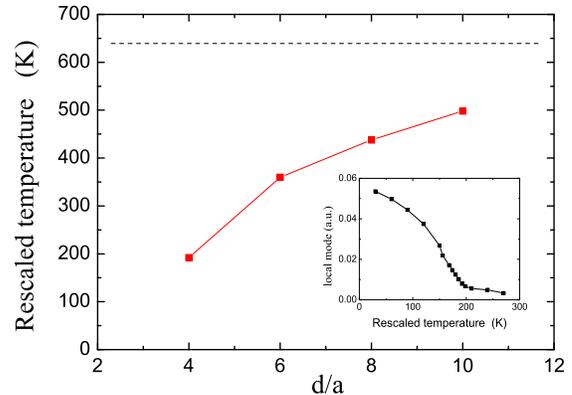}
\caption{Ferroelectric-paraelectric phase transition temperature
as the function of the film thickness $d$ under 2$\%$ compressive
strain. Dash line corresponds to $T_c$ of the bulk material.
Inset: the polarization amplitude as a function of the temperature
for $d=4$.}
\end{figure}

The stripe period for the film with $d=4$ (Fig.~2) is determined
to be $7.1\pm 0.5$ cell units ($\simeq 2.8 \pm 0.2$ nm) according
to the fact that 17 periodic stripe domains are observed in an
$120\times 5\times 4$ supercell. We find that the period $\lambda$
of the strip domain has the same order of the film thickness $d$
and increases with film thickness as shown in Fig.~4.

\begin{figure}[tbp]
\includegraphics[width=8.5cm]{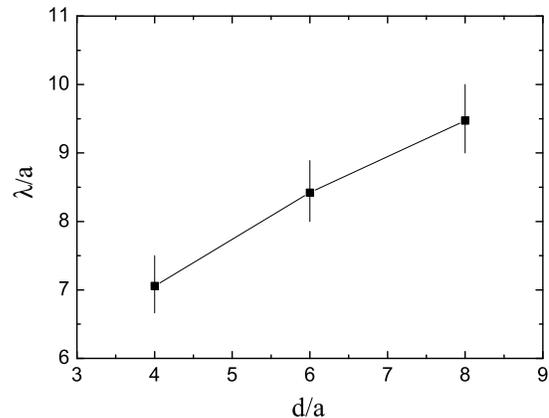}
\caption{The effect of the film thickness on the stripe period
under 2$\%$ compressive strain. The simulations for the period of
stripe domain are conducted respectively for $120 \times 5\times
4$, $160 \times 5\times 6$, and $180 \times 5\times 8$ supercells.
The errors are also indicated in the figure.}
\end{figure}

Our predictions are helpful for understanding many experimental
results. The stripe domains revealed here, with out-of-plane
polarizations and 180$^\circ$ domain walls extending through the
film, are essentially identical to what have been observed in PT
films by Streiffer \textit{et al.}~through the X-ray scattering
measurements\cite{str}. Their observed stripe period and the
dependence on the film thickness are consistent with our
calculated results. Based on our results, it is easy to understand
why no detectable polarization was probed by the electrostatic
force microscopy along the perpendicular directions of PZT
film\cite{ahn}, since it is demonstrated that the net polarization
vanishes by forming the stripe domain with very short period.

From a fundamental point of view, ferroelectricity in materials is
determined by the competition between the long-range electrostatic
and the short range covalent interactions. Our previous work
revealed that the orientation dependence of dipole-dipole
interaction is responsible for the appearance of unexpected phases
in compositional modulation PZT alloys near the morphotropic phase
boundary\cite{Hnd}. Here we indicate that the existence of the
critical thickness and periodic stripe domains can be also
understood by considering the orientation dependence of
dipole-dipole interaction. The long-range polarization
interactions are described in dipole-dipole interaction energy
\cite{WZ,LB2}:
\begin{equation}
E_{\mathrm{dipole}}=\frac{Z^{\ast 2}}{\epsilon _{\infty }}\sum_{i<j}\frac{%
\mathbf{u}_{i}\cdot \mathbf{u}_{j}-3(\mathbf{\hat{R}}_{ij}\cdot \mathbf{u}%
_{i})(\mathbf{\hat{R}}_{ij}\cdot \mathbf{u}_{j})}{R_{ij}^{3}},
\end{equation}%
where $Z^{\ast }$ is the Born effective charge and $\epsilon
_{\infty }$ the optical dielectric constant of the material. To
simplify the theoretical analyzes, we assume that all
$\mathbf{u}_{i}$ are along the $z$ axis. For the monolayer film,
the dipole-dipole interactions are constrained to be intralayered.
Then the second term of Eq. (1) is zero and the dipole-dipole
interaction coefficient is always positive for the ferroelectric
phase. Therefore, in the monolayer film, the dipole-dipole
interaction does not support the formation of ferroelectricity
along the $z$ axis. The ferroelectric phase will not appear in
this case unless the in-plane polarization is considered, as what
we have demonstrated above.

In contrary to the intralayer dipole interaction, the interlayer
dipole interaction is in favor of the ferroelectric order, and its
effect becomes more and more important with increasing film
thickness. So after the film thickness reaches a certain critical
value, the ferroelectricity appears. If the depolarization field
has not been eliminated by such effects as the surface charge
screening, domain structures will occur to decrease the intralayer
dipole interaction energy.

The variation of the stripe period with the film thickness comes
from the delicate balance between the long-range dipole
interaction and the short-range interaction. The dipole-dipole
interaction energy increases with increasing modulation period,
while the short-range interaction energy decreases with increasing
period. Compared with the short-range interaction, the dipole
interaction is long-range and thus more closely related to the
film thickness. This feature determines that the stripe period
varies with the film thickness. To further clarify this point, we
present a simple numerical analysis in the following. The local
modes are assumed to vary along the $x$ axis as
\begin{equation}
\mathbf{u_z}=u_{z0}\cos(\frac{2\pi x}{\lambda}) ,
\end{equation}%
where $\lambda$ is the period of the strip domain and $u_{z0}$ is
the amplitude of local mode $u_z$. In Fig.~5, the average
dipole-dipole interaction energy and short-long interaction energy
of one unit cell are shown as functions of $\lambda$ for thin
films with different thickness. The larger the period is, the more
quickly the dipole-dipole interaction energy decreases with
increasing film thickness. On the contrary, the energy difference
between different modulation periods is almost independent of the
film thickness for the short-range interaction, i.e., increasing
the film thickness only shifts the curve upward. Therefore, the
minimum of the total energy (see the inset in Fig.~5) moves to
higher $\lambda$ values while increasing the film thickness.

\begin{figure}[tbp]
\includegraphics[width=8.5cm]{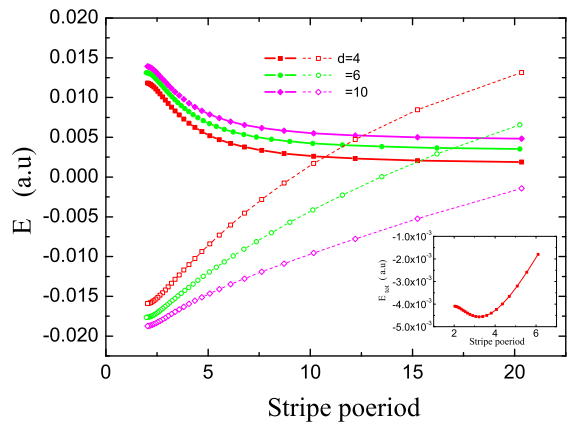}
\caption{The dipole-dipole interaction energy (dash line) and
short-range interaction energy (solid line) as functions of the
stripe period $\lambda$ for the thin film with different thickness
$d$. Inset: The sum of the dipole-dipole interaction energy and
short-range interaction energy as functions of the stripe period
for $d=4$.}
\end{figure}

The formation of domain structure helps to decrease the dipole
interaction energy by screening the depolarizing electrostatic
field. However, whether the polarization will appear is also
affected by the short-range interaction. If the minimum of the
total energy of domain structure is higher than that of the
paraelectric phase, the ferroelectricity will be suppressed, which
is just the cases occurring below the critical film thickness
(note that the total energy shown in Fig.~5 will increase with
decreasing thickness). It suggests that the critical thickness
exists even if the 180$^\circ$ domains are formed to screen the
depolarizing field.

In summary, we have demonstrated the existence of periodic
180$^\circ$ stripe domains with the out-of-plane polarization and
a critical thickness for ferroelectricity in disorder PZT thin
films, which may widely exist in perovskite thin films. It seems
that the electrical boundary conditions have little effect on the
critical thickness since the critical thickness is almost the same
for completely different electrical boundary condition. It is
pointed out that the critical thickness appears only when the
strain imposed by the substrate is strong enough to suppress the
in-plane polarization. Otherwise, ferroelectricity can be
maintained even in monolayer film.

This work was supported by State Key Program of Basic Research
Development of China (Grant No. TG2000067108), the National
Natural Science Foundation of China (Grant No. 10325415),
Trans-century Training Programme Foundation for the Talents by the
Ministry of Education of China, and China Postdoctoral Science
Foundation.


\begin{thebibliography}{}
\bibitem{note1} Although the above effective Hamiltonian was developed for
bulk PZT, the surface effect is partially and implicitly
considered during the simulations on the films by  i) including
the effect of the broken bonds in the surfaces, which is
substaintial for the surface energy; and by ii) relaxing the local
soft mode vectors and inhomogeneous strain, which partially
refelects the effect of surface relaxation.
\end{thebibliography}
\end{document}